\documentclass[preprintnumbers,amsmath,amssymb]{revtex4}

\usepackage{graphicx}
\usepackage{dcolumn}
\usepackage{bm}
\usepackage[hang]{subfigure}

\begin{document}
\title{In-situ frequency tuning of photons stored in a high Q microwave cavity.}
\author{M. Sandberg}
\email{martins@chalmers.se}
\affiliation{Department of Microtechnology and Nanoscience (MC2), Chalmers University of Technology, SE-412 96 G\"oteborg, Sweden}
\author{C. M. Wilson}
\affiliation{Department of Microtechnology and Nanoscience (MC2), Chalmers University of Technology, SE-412 96 G\"oteborg, Sweden}
\author{F. Persson}
\affiliation{Department of Microtechnology and Nanoscience (MC2), Chalmers University of Technology, SE-412 96 G\"oteborg, Sweden}
\author{G. Johansson}
\affiliation{Department of Microtechnology and Nanoscience (MC2), Chalmers University of Technology, SE-412 96 G\"oteborg, Sweden}
\author{V. Shumeiko}
\affiliation{Department of Microtechnology and Nanoscience (MC2), Chalmers University of Technology, SE-412 96 G\"oteborg, Sweden}
\author{T. Duty}
\affiliation{School of Physical Sciences, The University of Queensland, Brisbane QLD 4072 Australia.}
\author{P. Delsing}
\affiliation{Department of Microtechnology and Nanoscience (MC2), Chalmers University of Technology, SE-412 96 G\"oteborg, Sweden}
\date{\today}
\maketitle

\bf
Photons are fundamental excitations of the electromagnetic field and can be captured in cavities. For a given cavity with a certain size, the fundamental mode has a fixed frequency {\it f} which gives the photons a specific "colour". The cavity also has a typical lifetime $\tau$, which results in a finite linewidth $\delta${\it f}. If the size of the cavity is changed fast compared to $\tau$, and so  that the frequency change $\Delta${\it f} $\gg \delta${\it f}, then it is possible to change the "colour" of the captured photons.
Here we demonstrate superconducting microwave cavities, with tunable effective lengths. The tuning is obtained by varying a Josephson inductance at one end of the cavity. We show tuning by several hundred linewidths in a time $\Delta t \ll \tau$.  Working in the few photon limit, we show that photons stored in the cavity at one frequency will leak out from the cavity with the new frequency after the detuning. The characteristics of the measured devices make them suitable for dynamic coupling of qubits.\rm
\\

Superconducting transmission line resonators are useful in a number of applications ranging from X-ray photon detectors \cite{Day} to parametric amplifiers \cite{Haviland_parametric} and quantum
computation applications \cite {Wallraff_Nature, Majer_2007, sillanpaa}.  Very recently, there has been a lot of interest in tunable superconducting resonators \cite{Osborn, Lehnert_APL, saclay_tune}.  In these experiments the inductive properties of the Josephson junction is implemented as a tunable element and is tuned by a bias current or a magnetic field.
These devices have both large tuning ranges and high quality factors, however the speed at which these devices can be tuned has not been measured.

\begin{figure}[htb]
\subfigure{\includegraphics[width=9cm]{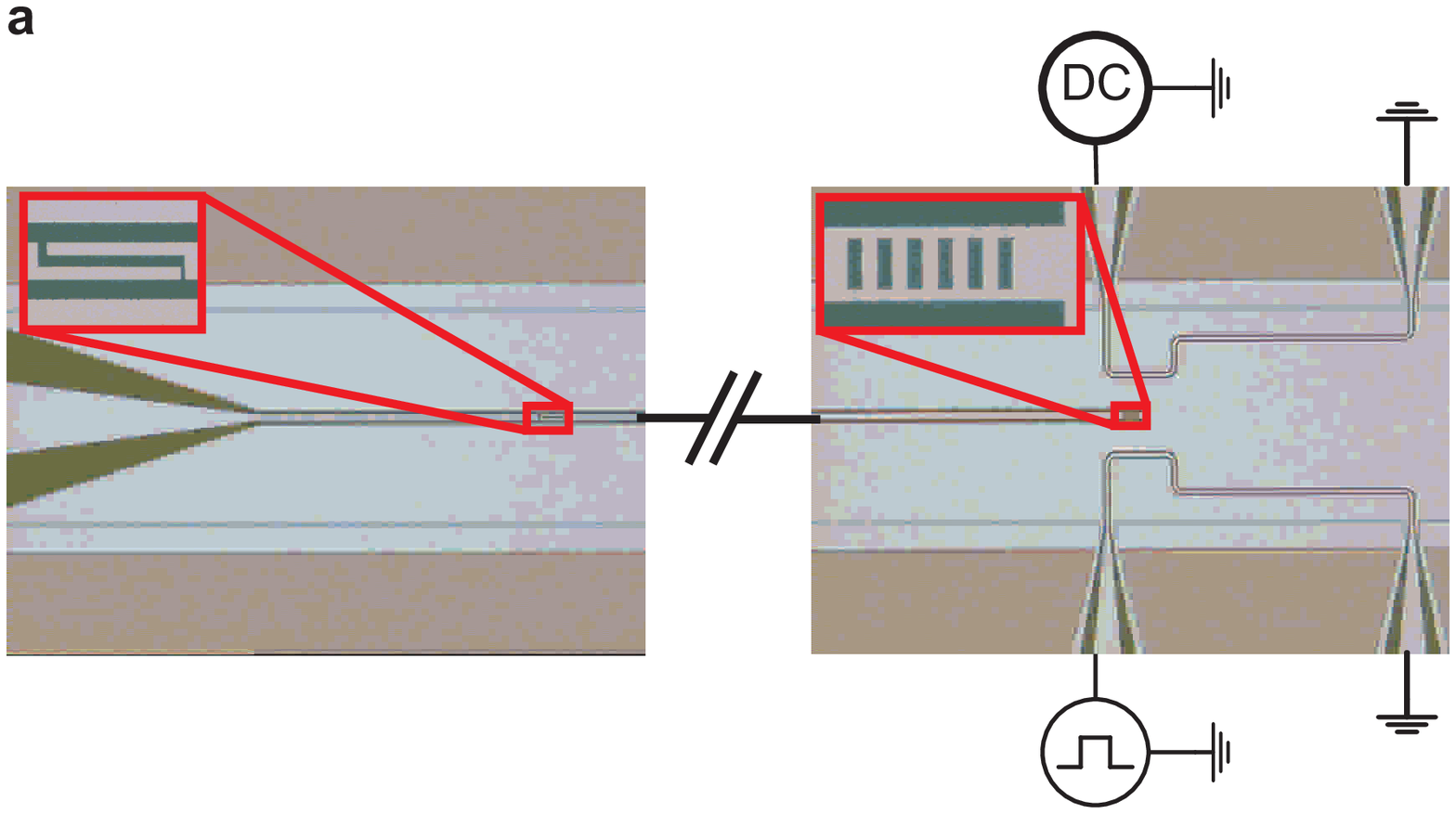}\label{fig:SQUIDSA}}
\subfigure{\includegraphics[width=9cm]{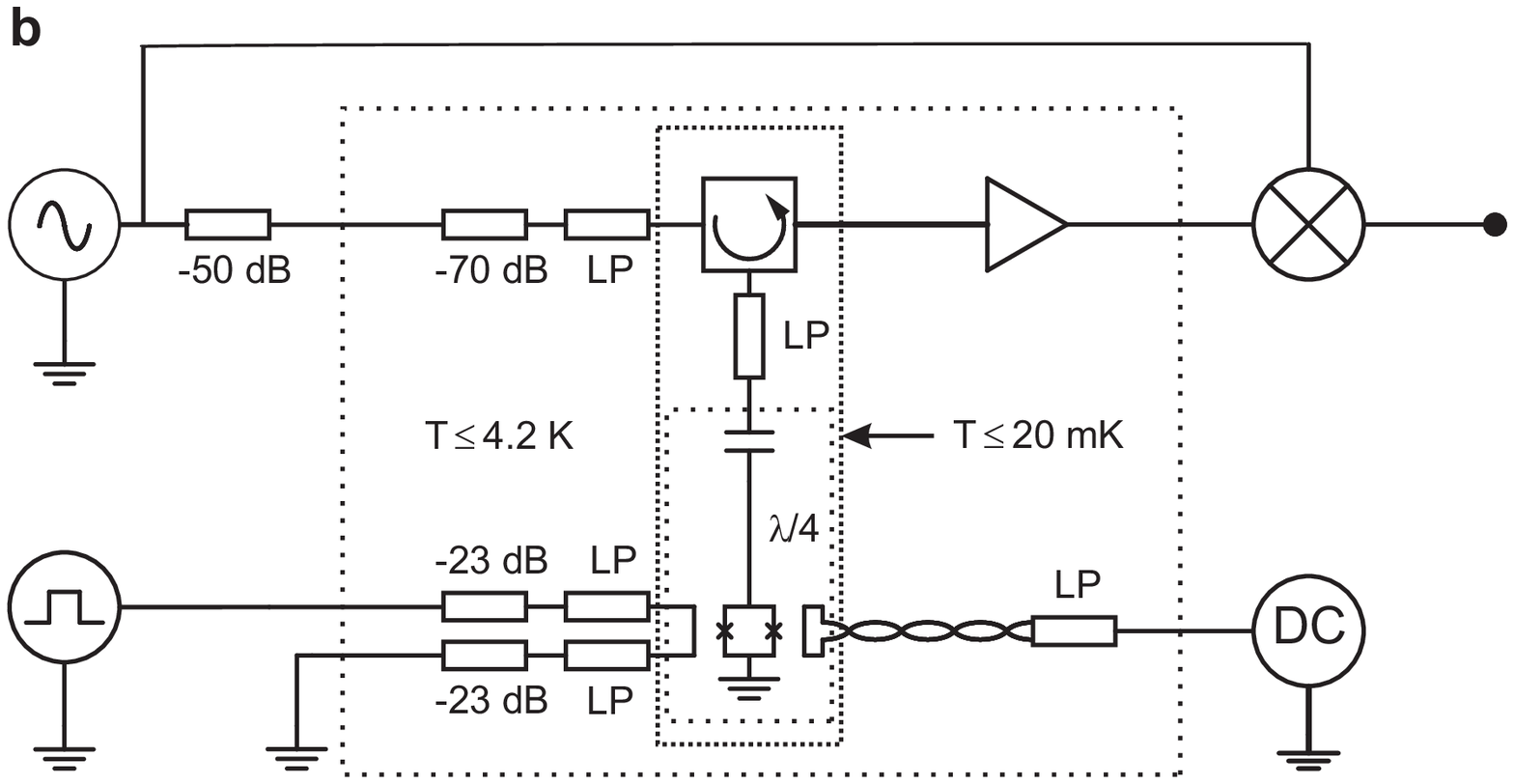}\label{fig:SQUIDSB}}
\caption{\bf Tunable resonator and measurement setup. \rm
\bf a, \rm  A micrograph of a typical tunable $\lambda/4$ resonator used in
the experiment.  The devices are fabricated on a silicon substrate
with a 400 nm layer of wet-grown silicon dioxide.  The cavity and the
Josephson junctions forming the SQUIDs are fabricated from aluminium
in a single lithography step using double angle evaporation and
oxidation.  The center strip is 13 $\mu$m wide and the gaps are 7
$\mu$m, giving a capacitance of 170 pF/m \cite{CAD} and a
characteristic impedance of 50 $\Omega$.
Two devices, A and B, were measured.  Device B had two on-chip flux
lines to tune the SQUIDs, one for DC bias and one for fast pulses,
while device A could only be tuned with an external magnetic field.
\bf b, \rm  Schematic measurement setup.  The samples are shielded from
external magnetic fields using an outer Pb shield and an inner cryoperm
shield.  The devices are probed through a circulator placed at the
base temperature of the dilution refrigerator and a HEMT amplifier
with a noise temperature of 4.5 K \cite{miteq} located at the 4 K
stage.  The excitation line has 70 dB of cold attentuation before the
input port of the circulator.  The fast flux line is a 50 $\Omega$
through-line going down to the chip and back to room temperature with
23 dB of attenuation in both directions.  The fast flux line is
filtered in both directions with 8 GHz low pass filters at the 600 mK
level.  The DC line has a 100 MHz low pass filter placed at 4 K. To do
fast tuning measurements, the signal from the resonator is mixed with
the drive signal.}
\end{figure}

The interaction between a qubit and
superconducting coplanar waveguide (CPW) resonator can, due to the small mode volume, be very strong when they are resonant with each other \cite{Blais_PRA}.  However, the
interaction can be modulated, becoming weak when the qubit and the
cavity are off-resonance.  Taking this into account, a protocol for a
controlled phase gate using two superconducting qubits and a tunable
resonator was constructed by Wallquist {\it et  al.}
\cite{Wallquist_PRB}.  The advantage of using a tunable resonator for
qubit interaction compared to a fixed resonator and tunable qubits is
that the qubits can stay at their optimal points
\cite{Vion_Quantronium} during operation.  Although a qubit design
has been presented that is insensitive to charge noise
\cite{Transmon}, there is still an optimal point in the flux direction
which may be essential to obtaining long coherence times.  To
implement the protocol suggested by Wallquist \emph{et al.}, a CPW
resonator with a large tuning range, fast tuning and high quality
factor is needed.  We have fabricated a quarter wavelength
($\lambda/4$) CPW resonator terminated to ground via one or several
Superconducting Quantum Interference Devices (SQUIDs) in series (see
Fig.  \ref{fig:SQUIDSA}).  In the other end of the resonator, a small
coupling capacitance is placed through which the resonator can be
excited and probed using microwave reflectometry.

A SQUID can be viewed as a lumped element inductor with the nonlinear
inductance

\begin{equation}
\label{eq:Ls}
L_s=\frac{\Phi_0}{4 \pi I_c|\cos(\pi\Phi/\Phi_0)|\sqrt{1-\left(\frac{I}{2I_c\cos(\pi\Phi/\Phi_0)}\right)^2}}
\end{equation}

where $\Phi$ is the applied magnetic flux, $\Phi_0=\hbar/2e$ is the
magnetic flux quantum, $I_c$ is the critical current of each SQUID
junction and $I$ is the current through the SQUID. The inductance can
be varied by applying a magnetic field or a bias current.  When the
current $I$ is much smaller than the suppressed critical current,
$L_s$ can be considered linear.  A real SQUID has also a capacitance
$C_s$ and the subgap resistance $R_s$ in parallel with the inductance
so that the total impedance of the SQUIDs is

\begin{equation}
\label{eq:Zs}
Z_s=N\left(\frac{i\omega L_s R_s}{R_s+i\omega L_s -\omega^2L_s C_s R_s}\right)
\end{equation}

where $N$ is the number of SQUIDs in series.

The resonator circuit can be described using the scattering matrices
formulation \cite{Collin} using three scattering matrixes in series.
When probed with a probe voltage signal $V_d$, we find the reflection
coefficient

\begin{equation}
\label{eq:S11}
\Gamma=\frac{V_r}{V_d}=\left( S_{11}+\frac{S_{12}S_{21}\Gamma_s e^{-2\gamma l}}{1-S_{22}\Gamma_s e^{-2\gamma l}}\right).
\end{equation}
\\

where the $S_{ij}$ are the scattering parameters for the coupling
capacitance, $l$ is the length of the cavity, $\gamma = \alpha +i\beta$ is the complex wave number of
the transmission line, $\Gamma_s$ is the reflection
coefficient of the SQUIDs, obtained using equation  \ref{eq:Zs} as a load
impedance.  The parameter $\alpha$ is the attenuation along the transmission
line and for a non-tunable $\lambda/4$ resonator $\alpha = \pi/\lambda
Q_{int}$, where $Q_{int}$ is the internal quality factor of the
device.  $\beta = 2\pi/\lambda$, is the phase propagation of the transmission line.

Measurements of $\Gamma$ on two different devices are presented here.
Device A has one SQUID and is tuned by an external magnetic field,
while device B has six SQUIDs in series and is tuned by an on-chip flux bias
(see Table \ref{tab:A_B}).  A micrograph of a device identical to
device B is showed in Fig.  \ref{fig:SQUIDSA}.  Measurements of the
devices were performed at the base temperature (below 20 mK) of a
dilution cryostat.  The measurement setup is shown in Fig.
\ref{fig:SQUIDSB}.

\begin{figure}
\subfigure{\includegraphics[width=7cm]{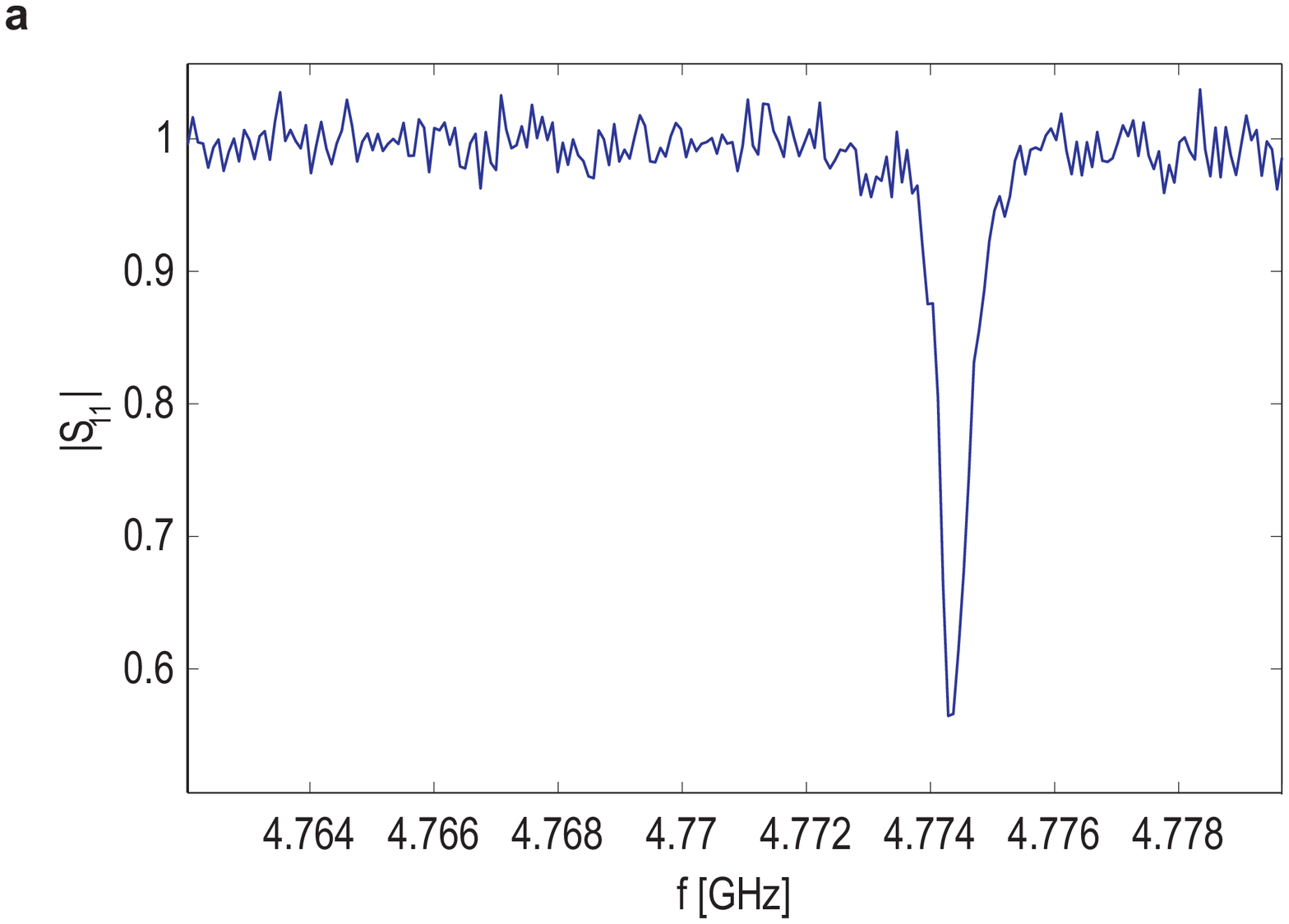}\label{fig:res}}
\subfigure{\includegraphics[width=7cm]{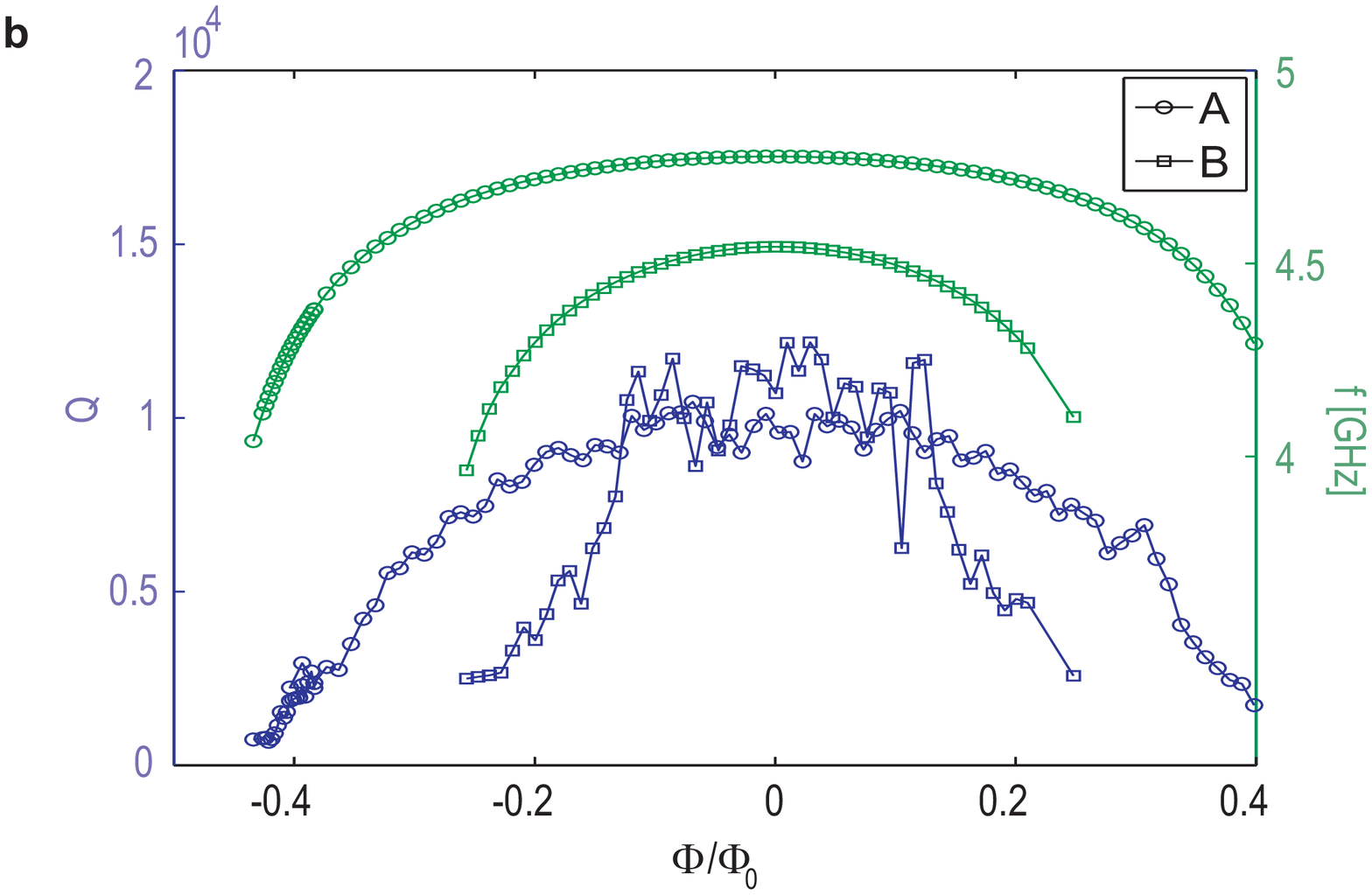}\label{fig:DC}}
\caption{\bf Microwave reflectometry measurements. \rm \bf a, \rm Magnitude of the reflection spectrum of sample A.  The resonance frequency is given by the center of the dip
in the spectrum and the Q value is the resonance frequency divided by
the full width at half maximum.  From these measurements, we can
conclude that the devices are undercoupled \emph{i.e.} the dominate
source of loss is from inside the cavity and not from leakage through
the coupling capacitance.  The internal losses have four
contributions: radiation, resistive losses in the conductor,
dielectric losses, and losses in the SQUID. In the CPW design, the
radiation losses are small.  The resistive losses should also be small
since the temperature of the device is well below the critical
temperature of aluminium and the applied frequency is well below the
superconducting gap frequency.  The losses must therefore either be
due to dielectric losses or due to the SQUIDs.  To address this
question, a non-tunable reference device (without SQUIDs) was
fabricated and measured, not shown here.  The measurements showed
undercoupled behaviour for this device at low drive powers, suggesting
that the SQUIDs are not the limiting factor.  As we increased the
drive power however, the $Q$ of the reference devices increased and we
observed a transition from undercoupled to overcoupled behaviour.  The
power dependence seems to be consistent with what Martinis \emph{et
al.} observed for losses due to two-level fluctuators in the
dielectric \cite{Martinis_PRL_2005}.  \bf b, \rm  Resonance frequency (green)
and Q value (blue) as a function of applied magnetic flux for the two
devices A (circles) and B (squares).  Device A has one SQUID with a
critical current of $I_c\approx 1.2$ $\mu$A. Device B has six SQUIDSs
in series, each with a critical current of $I_c\approx 2.3$ $\mu$A.}
\end{figure}

\begin{figure}
\begin{centering}
\subfigure{\includegraphics[width=7cm]{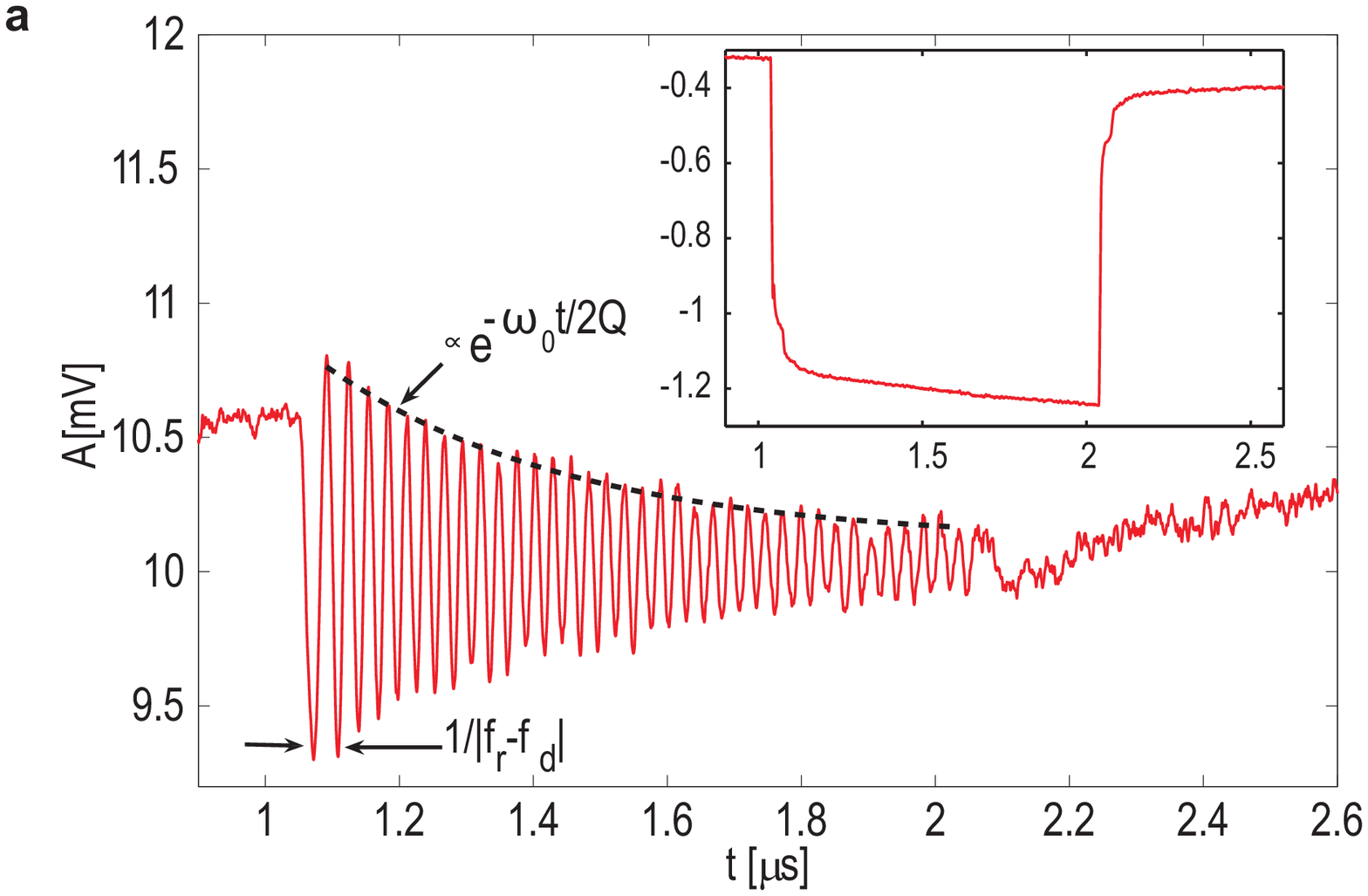}\label{fig:exp_a}}
\subfigure{\includegraphics[width=7cm]{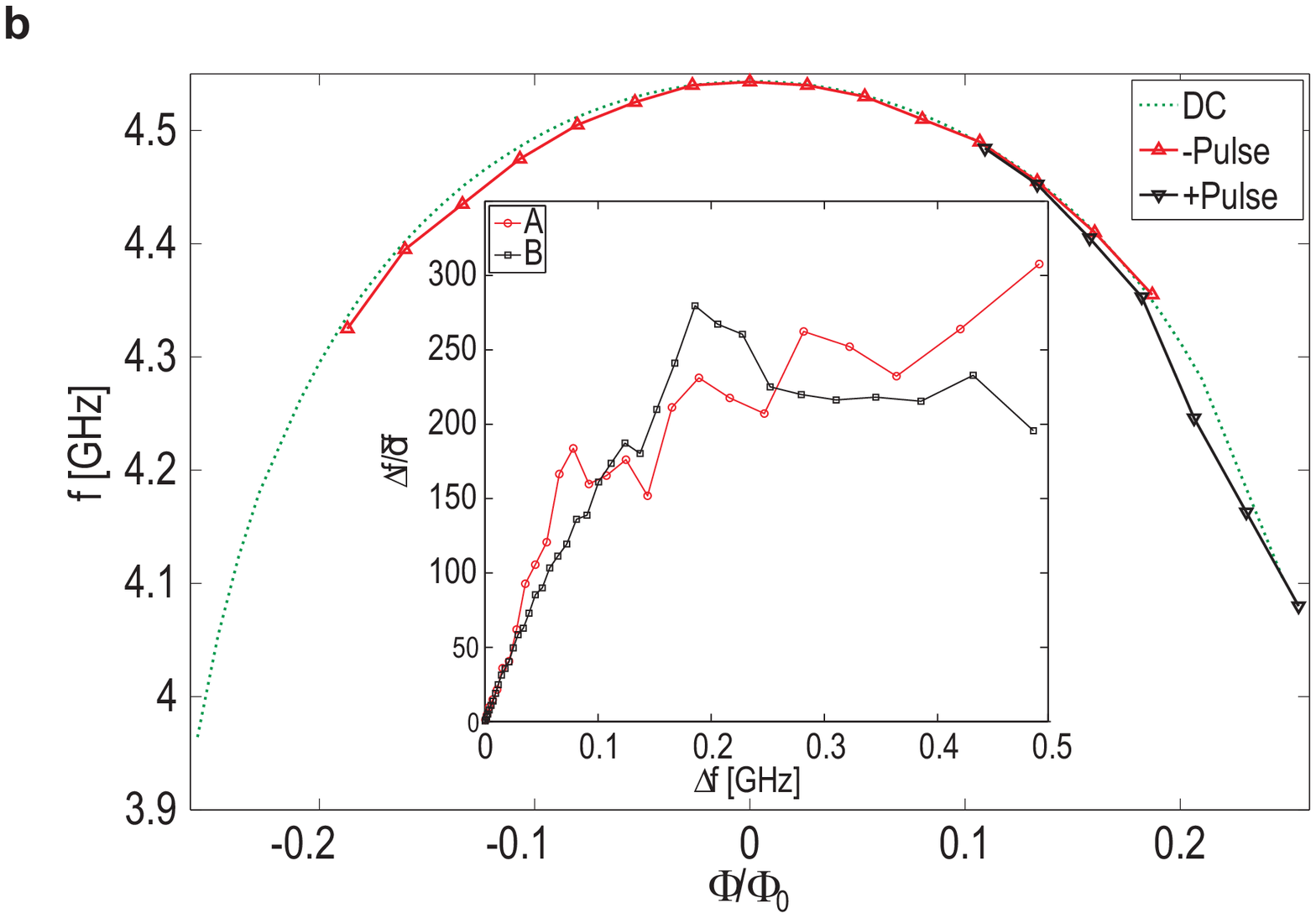}\label{fig:exp_c}}
\end{centering}
\caption{
\bf Fast detuning measurements. \rm
\bf a, \rm To measure the tuning speed, we first drive the resonator on
resonance until equilibrium is reached.  A fast flux pulse is then
applied to detune the resonator from the drive.  We then measure the
signal from the resonator as the stored energy decays.  Mixing the
signal with the drive frequency and filtering out the high frequency
components (see Fig.  \ref{fig:SQUIDSB} b)) we observe a decaying,
oscillating signal where the frequency of the oscillations is the
difference frequency of the drive and the detuned resonant frequency
of the resonator.  From the decay $\tau$ of the oscillations the $Q$
value can be obtained as $Q=\omega\tau/2$, where the factor of 2 is
due to the fact that amplitude and not power is measured. The traces are obtained by repeatedly applying a pulse and then average the signal over several pulses. A typical number of average used is between one hundred thousand and one million. The decay measurements showed a Q of almost a factor of two less than the value obtained from line width measurements. The discrepancy can be explained by the distribution in amplitude of the applied flux pulse. As the data is average over a large number of pulses a small distribution in pulse height cause an effective increase in the decay. Inset: the pulse used to detune the resonance frequency.  The amplitude of the pulse sets how far the
resonator is detuned.  \bf b, \rm  Resonance frequency obtained from DC
measurements compared to values obtained from fast pulse measurements.
We can both increase and decrease the frequency with the fast pulse
(indicated by up and down pointing triangles respectively) meaning
that we can both stretch and compress the photons in the cavity.  The
inset shows how many linewidths that the devices A and B are detuned
as a function of detuning.  }
\end{figure}

Measuring the reflection coefficient as a function of applied magnetic
DC field, Fig.  \ref{fig:DC}, using a vector network analyzer (VNA)
and fitting equation  \ref{eq:S11} we get $\alpha \sim 0.106$ dB/m and
$\beta \sim 241$ rad/m, on resonance, for sample A at zero detuning.
The value of $\beta$ on resonance gives a cavity electrical length corresponding to
86 degrees, compared to 90 degrees for a regular $\lambda/4$
resonator.  We observe a tunability $\Delta f_{max}$ of 700 MHz for
sample A and 480 MHz for sample B (see Fig.  \ref{fig:DC}).  The $Q$
value is obtained from the line width $\delta f$ and depends on the
detuning.  Therefore, the number of detuned line widths $\Delta
f/\delta f$ as a function of detuning gives a figure of merit for the
devices.  This is ploted in the inset of Fig.  \ref{fig:exp_c}.  The
maximum $\Delta f/\delta f$ is similar for the two devices and is
around 250.  We observe that the $Q$ decreases as the magnetic field
is applied (see Fig.  \ref{fig:DC}).  We do not yet understand this
quantitatively.  One possible reason could be inhomogenous broadening
due to flux noise, but the source of this noise is not known.  In
particular, the typical values of $1/f$ flux noise or $1/f$ critical
current noise are orders of magnitude smaller than needed to explain this effect.

\begin{table}
\begin{tabular}{c|c|c|c|c|c|c}
  Sample & N & I$_c$ & {\it f}$_0$ & $\Delta${\it f}$_{Max}$ & Q$_{0}$ & Flux bias \\ \hline
  A & 1 & 1.2 $\mu A$ & 4.77 GHz & 700 MHz & 10000 & External\\ \hline
  B & 6 & 2.3 $\mu A$ & 4.54 GHz & 480 MHz & 12000 & On chip\\
\end{tabular}
\caption{\label{tab:A_B} Parameters of the two measured devices.  {\it N} is
the number of SQUIDs, {\it I}$_c$ is the critical current of the Josephson
junctions, {\it f}$_0$ is the zero flux resonance frequency,
$\Delta${\it f}$_{Max}$ is the maximum detuning achieved, {\it Q}$_{0}$ is the
zero field {\it Q} value and the last column states how the flux bias was
applied to the SQUID.}
\end{table}


One of the most important aspects of the device for qubit coupling is
the tuning speed.  A measurement of the tuning speed that relies on a
change of the reflection coefficient will be limited by the ring up
time of the resonator $Q/\omega_0$.  In our case, this will be of the
order of microseconds.  This is however not the same as the speed at
which the resonator can be tuned.  If we apply fast flux pulses as we
drive the resonator on resonance and mix the signal from the resonator
with the drive we observe a beating between the drive frequency and
the frequency of the photons leaking out of the resonator.  This
demonstrates that we in fact can tune the cavity much faster then the
ring up time (see Fig.  \ref{fig:exp_a}).

By varying the amplitude of the flux pulse, we can map out the
resonance frequency as a function of flux bias, Fig.  \ref{fig:exp_c}.
We can apply positive and negative flux pulses and in that way we can
either increase or decrease the resonance frequency during the pulse.  The
leakage of energy at the detuned frequency suggests that we can both
stretch and compress the photons stored in the resonator
\emph{in-situ} as we apply the flux pulse.  The energy stored in the
cavity on resonance can be obtained as twice the average electrical
energy, and ignoring zero point fluctuations

\begin{equation}
\label{eq:energy}
E=\hbar\omega_0N_p=2 \int^{l}_{0} \frac{c_lV(x)V^{\ast}(x)}{2}dx
\end{equation}

where $V(x)$ is voltage along the transmission line and $c_l$ is the
capacitance per unit length.
For the fast tuning experiment we apply a power of -141 dBm (assuming
5 dB loss in the cables) giving an average drive photon number $N_p$
in the resonator of about 5 photons. The average number of thermal photons is much smaller than one since the temperature of the cavity is approximately 20 mK, whereas the cavity frequency corresponds to $\sim$ 225 mK. The low photon number suggest
that we in fact change the frequency of individual photons as we tune
the resonator with a fast flux pulse. As the cavity is detuned, the photons
inside the cavity are forced to adjust to the new resonance frequency.
When compressing the photons our external magnetic field is thus doing work on the photons increasing their energy.

\begin{figure}
\begin{centering}
\includegraphics[width=7cm]{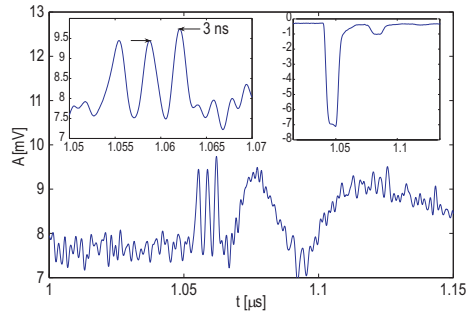}
\end{centering}
\caption{\label{fig:ns_pulse}\bf Tuning the cavity in $\le$ 10 ns. \rm
Signal when a 10 ns pulse is applied.  The rigth inset shows the
applied pulse.  The detuning is 330 MHz, seen from the period of the
oscillations (left inset).  Imperfections in the bias line causes
small multiple reflections of the pulse as seen in the rigth inset,
causing the features seen after the fast oscillations in the main
figure.}
\end{figure}

We can obtain an estimate on how fast we can tune the device by
decreasing the duration time of the flux pulse.  If we apply a large
detuning pulse, the observed oscillations are fast, allowing us to
apply a short pulse and still observe the oscillations.  In this way,
we can still observe oscillations even decreasing the duration of the
pulse down to a few ns.  In Fig.  \ref{fig:ns_pulse} a detuning of
330 MHz for a 10 ns pulse is shown.  The rise time of the flux pulse
is here only about 3 ns.  Due to imperfections in the fast pulse line,
we get reflections (see lower inset in Fig.  \ref{fig:ns_pulse})
causing a slower oscillation to occur after the fast oscillations.
The pulse can be decreased even further until only one oscillation is
observed (not showed here), but the duration of the flux pulse is then
of the same order as the rise time.



In conclusion, we have designed and measured a tunable superconducting
CPW resonator.  We have demonstrated a tunability of 700 MHz for a 4.9
GHz device.  As a figure of merit, we see that we can detune the
devices more than 250 corrected linewidths.  We have also demonstrated
that our device can be tuned substantially faster than its decay time,
allowing us to change the frequency of the energy stored in the
cavity.  Having done this in the few photon limit, we therefore assert
that we can tune the frequency of individual microwave photons stored
in the cavity.  This can be done by several hundred MHz on the
timescale of nanoseconds.

We believe that the high $Q$ value, large tunability, and the fast
tuning make this device very suitable for dynamic coupling of qubits.
It should also be an interesting system for the study of fundamental
physics, such as generation of nonclassical photon states
\cite{Agarwal}.
\\

Acknowledgements\\
We acknowledge financial support from the Swedish VR and SSF, from the Wallenberg foundation and from the European Union under the EuroSQIP project.  We would also like to acknowledge fruitful discussions with M. Wallqusit, P. Bertet and D. Esteve.

\bibliographystyle{nature}

\end{document}